\documentclass[10pt, technote]{IEEEtran}

\usepackage[pdftex]{graphicx}
\DeclareGraphicsExtensions{.pdf,.jpeg,.png}

\usepackage{array}
\usepackage[cmex10]{amsmath}
\usepackage{url}
\usepackage{color}

\usepackage{subfigure}
\usepackage{esint} 
\usepackage{tabu}

\newcommand{\BE}{\begin{equation}}   
\newcommand{\EE}{\end{equation}}     
\newcommand{\BEn}{\begin{equation*}} 
\newcommand{\EEn}{\end{equation*}}   
\newcommand{\BF}{\begin{figure}}     
\newcommand{\EF}{\end{figure}}  	 
\newcommand{\BFw}{\begin{figure*}}   
\newcommand{\EFw}{\end{figure*}}     
\newcommand{\BT}{\begin{table}}		 
\newcommand{\ET}{\end{table}}        

\providecommand{\RE}{\ensuremath{\mathrm{Re}}} 
\providecommand{\IM}{\ensuremath{\mathrm{Im}}} 

\newcommand{\J}{\mathrm{j}}                 

\providecommand{\D}{\,\mathrm{d}}           
\providecommand{\V}[1]{\boldsymbol{#1}}     

\graphicspath{} 

\newcommand{\Wm}{\widetilde{W}_\mathrm{m}}
\newcommand{\We}{\widetilde{W}_\mathrm{e}}
\newcommand{\Wvan}{W^\mathrm{(Vandenbosch)}}

\begin{document}
\title{Comments on ``On Stored Energies and Radiation Q"}
\author{Miloslav~Capek,~\IEEEmembership{Member,~IEEE,}
        and~Lukas~Jelinek
\thanks{Manuscript received XXX, 2015; revised XXX, 2015.}
\thanks{The authors are with the Department of Electromagnetic Field, Faculty of Electrical Engineering, Czech Technical University in Prague, Technicka 2, 16627, Prague, Czech Republic
(e-mail: \mbox{miloslav.capek@fel.cvut.cz}, \mbox{lukas.jelinek@fel.cvut.cz}).}
}

\markboth{Journal of \LaTeX\ Class Files,~Vol.~6, No.~1, January~2007}%
{Capek and Jelinek: Comments to ``On Stored Energies and Radiation Q" by W. Geyi}
\maketitle

\IEEEPARstart{T}{he} commented paper \cite{Geyi_OnStoredEnergiesAndRadiationQ} claims to provide a new expression for an energy stored around a general radiator. The major purpose of this comment is to show that the claim is unjustified. Alongside with this issue, it is pointed out that some of the core formulas of \cite{Geyi_OnStoredEnergiesAndRadiationQ} are not completely correct, and that their correct form has in fact been derived elsewhere, though for the purpose of evaluating the quality factor $Q_Z$ and not the stored energies. 

The major outcome of the commented paper \cite{Geyi_OnStoredEnergiesAndRadiationQ} is claimed to be equations (29)--(31), which should represent the total stored energy \mbox{$\Wm+\We$}, the stored magnetic energy $\Wm$, and the stored electric energy $\We$ (using the notation of \cite{Geyi_OnStoredEnergiesAndRadiationQ}), and their connection with (2), (4), (5) of \cite{Geyi_OnStoredEnergiesAndRadiationQ}. Equations (4), (5) of \cite{Geyi_OnStoredEnergiesAndRadiationQ} are however not valid under the assumptions of \cite{Geyi_OnStoredEnergiesAndRadiationQ}, see \cite{AndersenBerntsen_CommentsOnTheFosterReactanceTheoremGeyi} and \cite{Best_TheFosterReactanceTheoremAndQualityFactorForAntennas}. 

In order to be more specific, we will first recall here a well-established result \cite{Rhodes_ObservableStoredEnergiesOfElectromagneticSystems}, \cite{YaghjianBest_ImpedanceBandwidthAndQOfAntennas}
\BE
\label{TAPcom_EQ1}
\begin{split}
\frac{\partial X}{\partial \omega } \left| {{I_0}} \right|^2 =& \int\limits_0^\infty  \int\limits_\Omega \left( {\mu {{\left\| {\V{H}} \right\|}^2} + \epsilon {{\left\| {\V{E}} \right\|}^2} - 2\epsilon \frac{{{{\left\| {\V{F}} \right\|}^2}}}{{{r^2}}}} \right){r^2}\,{\mathrm{d}}\Omega \,{\mathrm{d}}r \\
&+ 2 \sqrt{\frac{\epsilon}{\mu}} \, \IM \int\limits_\Omega {\left. {\frac{\partial \V{F}}{\partial\omega}} \right|_{I_0}} \cdot \V{F}^\ast\,\mathrm{d} \Omega,
\end{split}
\EE
where $X$ is the input reactance of the considered radiator, which is fed by a frequency independent\footnote{The assumption of the frequency independent current at the input port is essential for the validity of (\ref{TAPcom_EQ1}) and of (\ref{TAPcom_EQ6}), (\ref{TAPcom_EQ7}) later on. This assumption is not used in \cite{Geyi_OnStoredEnergiesAndRadiationQ}. However, as the system at hand is linear, this input current normalization can always be made. Any result claimed in \cite{Geyi_OnStoredEnergiesAndRadiationQ} as valid for general input current thus must be valid also for the frequency independent input current. It is also important to stress that the frequency independence of the input current does not imply the frequency independence of the current density on the entire radiator.} current $I_0$, and which is surrounded by an homogeneous, isotropic and non-dispersive medium of permittivity $\epsilon$ and permeability $\mu$. Equation (\ref{TAPcom_EQ1}) further assumes $r$ to be the radial coordinate, \mbox{${\mathrm{d}}\Omega  = \sin \theta \D \theta \D \varphi$} being the solid angle differential, $\V{E}$, $\V{H}$ being the electric and magnetic intensity generated by the antenna, respectively, and \mbox{${\V{F}}\left( {\V{r}} \right) = \lim_{r \to \infty } \left\{{r {\V{E}}\left( {\V{r}} \right){\mathrm{exp}\left(\J k r\right)}}\right\}$} with $k$ being the wave-number. The total radiated power can be written as \cite{Rhodes_ObservableStoredEnergiesOfElectromagneticSystems}, \cite{YaghjianBest_ImpedanceBandwidthAndQOfAntennas}
\BE
\label{TAPcom_EQ2}
P_\mathrm{rad} = \frac{1}{2} \sqrt{\frac{\epsilon}{\mu}} \int\limits_\Omega \left\| \V{F} \right\|^2\,\mathrm{d}\Omega.
\EE
It follows from (2) that (2) of \cite{Geyi_OnStoredEnergiesAndRadiationQ} can be rewritten as
\BE
\label{TAPcom_EQ3}
\begin{split}
\Wm &+ \We = \lim\limits_{r\to\infty} \left(W_\mathrm{m} + W_\mathrm{e} - \frac{r}{\nu} P_\mathrm{rad} \right)= \\
&= \frac{1}{4} \int\limits_0^\infty  \int\limits_\Omega \left( {\mu {{\left\| {\V{H}} \right\|}^2} + \epsilon {{\left\| {\V{E}} \right\|}^2} - 2\epsilon \frac{{{{\left\| {\V{F}} \right\|}^2}}}{{{r^2}}}} \right){r^2}\D\Omega\D r,
\end{split}
\EE
in which $\nu = 1 / \sqrt{\mu\epsilon}$. Equation (\ref{TAPcom_EQ3}) together with (4), (5) of \cite{Geyi_OnStoredEnergiesAndRadiationQ} thus results in
\BE
\label{TAPcom_EQ4}
\frac{\partial X}{\partial \omega } \left| {{I_0}} \right|^2 \stackrel{?}{=} \int\limits_0^\infty  \int\limits_\Omega \left( {\mu {{\left\| {\V{H}} \right\|}^2} + \epsilon {{\left\| {\V{E}} \right\|}^2} - 2\epsilon \frac{{{{\left\| {\V{F}} \right\|}^2}}}{{{r^2}}}} \right){r^2} \D \Omega \D r,
\EE
which is not generally correct as it lacks the last term on the RHS of (\ref{TAPcom_EQ1}), which is generally non-zero \cite[Fig.~10]{YaghjianBest_ImpedanceBandwidthAndQOfAntennas}. This omission has in fact already been mentioned in \cite[text above Eq. (72)]{YaghjianBest_ImpedanceBandwidthAndQOfAntennas} and \cite[text above Eq. (4)]{Best_TheFosterReactanceTheoremAndQualityFactorForAntennas}, but has been missed by the author of \cite{Geyi_OnStoredEnergiesAndRadiationQ} despite of \cite{YaghjianBest_ImpedanceBandwidthAndQOfAntennas} being cited in \cite{Geyi_OnStoredEnergiesAndRadiationQ}.

Alongside with the above argumentation, it is worth noting that the questionable association (result of (4), (5) of \cite{Geyi_OnStoredEnergiesAndRadiationQ})
\BE
\label{TAPcom_EQ5}
\displaystyle \frac{1}{4}\frac{{\partial {X}}}{{\partial \omega }} \left| I_0 \right|^2 \stackrel{?}{=} \Wm + \We, 
\EE
can be tested directly. To this point, the solid line in Figs.~\ref{fig1},~\ref{fig2} shows that \mbox{$\partial X / \partial\omega$} can be negative for radiating systems, unlike to the stored electromagnetic energy. This well established theoretical and experimental fact (see for example \cite{Best_TheFosterReactanceTheoremAndQualityFactorForAntennas}) is omitted by the author of \cite{Geyi_OnStoredEnergiesAndRadiationQ}. 

The next commentary concerns the seemingly erroneous relation of (29)--(31) of \cite{Geyi_OnStoredEnergiesAndRadiationQ} with (4), (5) of \cite{Geyi_OnStoredEnergiesAndRadiationQ}, which can be written as
\BE
\label{TAPcom_EQG1}
\begin{array}{l}
\displaystyle \frac{1}{4}\frac{{\partial {X}}}{{\partial \omega }} \left| I_0 \right|^2 \stackrel{?}{=} \Wvan  \\
\displaystyle - \frac{{\omega \eta \nu }}{{{8\pi }}}\int\limits_{V_1} {\int\limits_{V_2}  { \IM \left\{ {{\rho^\ast}\left( {{\V{r}_1}} \right)\frac{{\partial \rho \left( {\V{r}_2} \right)}}{{\partial \omega }}} \right\}\frac{{\sin\left( {kR} \right)}}{R}\D{V_2}} \D{V_1}} \\
\displaystyle + \frac{{\omega \eta \nu }}{{{8\pi }}}\int\limits_{V_1} {\int\limits_{V_2}  {\frac{1}{{{\nu^2}}} \IM \left\{ {{{\V{J}}^\ast}\left( {{\V{r}_1}} \right) \cdot \frac{{\partial {\V{J}}\left( {\V{r}_2} \right)}}{{\partial \omega }}} \right\}\frac{{{\sin}\left( {kR} \right)}}{R}\D{V_2}} \D{V_1}}, 
\end{array}
\EE
where
\BE
\label{TAPcom_EQG2}
\begin{array}{l}
\Wvan = \\
\displaystyle  \frac{{\eta \nu }}{{16{\pi }}}\int\limits_{V_1} {\int\limits_{V_2}  {\left( {\rho^\ast \left( {\V{r}_1} \right){\rho}\left( {{\V{r}}_2} \right) + \frac{1}{{{\nu ^2}}}{\V{J}^\ast}\left( {\V{r}_1} \right) \cdot {{\V{J}}}\left( {{\V{r}_2}} \right)} \right)\frac{{{\cos}\left( {kR} \right)}}{R}\D{V_2}} \D{V_1}} \\
\displaystyle + \frac{{\omega \eta }}{{16{\pi }}}\int\limits_{V_1} {\int\limits_{V_2}  {\left( {\rho^\ast \left( {\V{r}_1} \right){\rho}\left( {{\V{r}_2}} \right) - \frac{1}{{{\nu^2}}}{\V{J}^\ast}\left( {\V{r}_1} \right) \cdot {{\V{J}}}\left( {{\V{r}_2}} \right)} \right)\sin \left( {kR} \right)\D{V_2}}\D{V_1}}
\end{array}
\EE
is an energy defined in \cite[Eq.~(63),~(64)]{Vandenbosch_ReactiveEnergiesImpedanceAndQFactorOfRadiatingStructures}, ${\rho \left( {\V{r}} \right)}$ is the charge density, and $\eta$ is the wave impedance. In fact, the correct version of (\ref{TAPcom_EQG1}) has previously been derived in \cite{CapekJelinekHazdraEichler_MeasurableQ,CapekJelinekHazdraEichler_MeasurableQ_Arxive} and reads (using the notation of this comment)
\BE
\label{TAPcom_EQ6}
\begin{array}{l}
\displaystyle \frac{1}{4}\frac{{\partial {X}}}{{\partial \omega }} \left| I_0 \right|^2 = \Wvan  \\
\displaystyle - \frac{{\omega \eta \nu }}{{{8\pi }}}\int\limits_{V_1} {\int\limits_{V_2}  { \RE \left\{ {\frac{1}{\omega }{\rho^\ast}\left( {{\V{r}_1}} \right)\frac{{\partial \omega \rho \left( {\V{r}_2} \right)}}{{\partial \omega }}} \right\}\frac{{\cos\left( {kR} \right)}}{R}\D{V_2}} \D{V_1}} \\
\displaystyle + \frac{{\omega \eta \nu }}{{{8\pi }}}\int\limits_{V_1} {\int\limits_{V_2}  {\frac{1}{{{\nu^2}}} \RE \left\{ {{{\V{J}}^\ast}\left( {{\V{r}_1}} \right) \cdot \frac{{\partial {\V{J}}\left( {\V{r}_2} \right)}}{{\partial \omega }}} \right\}\frac{{{\cos}\left( {kR} \right)}}{R}\D{V_2}} \D{V_1}},
\end{array}
\EE
where the input current $I_0$ is assumed to be frequency independent. Formula (\ref{TAPcom_EQG1}) differs from (\ref{TAPcom_EQ6}) in the last two terms. The Figs.~\ref{fig1},~\ref{fig2} show which of the two formulas better represents the frequency change of the input reactance calculated as a ratio between voltage and current at the input port. The comparison is performed on an example of thin dipole similar to that of \cite{Geyi_OnStoredEnergiesAndRadiationQ} and on an example of Yagi-Uda antenna treated in \cite{YaghjianBest_ImpedanceBandwidthAndQOfAntennas}. In order to further support our claims, an additional and independent evaluation of \mbox{$\partial X / \partial\omega$} can be found. This evaluation relies on a combination of (\ref{TAPcom_EQ1}) and results of \cite[Eqs.~(4),~(5)]{GustafssonJonsson_AntennaQandStoredEnergiesFieldsCurrentsInputImpedance} or similarly \cite[Eqs.~(25),~(26),~(28)]{Gustaffson_StoredElectromagneticEnergy_PIER}, which reads
\BE
\label{TAPcom_EQ7}
\begin{array}{l}
\displaystyle \frac{1}{4}\frac{{\partial {X}}}{{\partial \omega }} \left| I_0 \right|^2 = \Wvan  \\
\displaystyle - \frac{{\eta \nu }}{{{16\pi }}}\int\limits_{V_1} {\int\limits_{V_2}  { \IM \left\{ {{\rho^\ast}\left( {{\V{r}_1}} \right){\rho \left( {\V{r}_2} \right)}} \right\} \mathcal{G} \left(\boldsymbol{r}_1, \boldsymbol{r}_2\right) \D{V_2}} \D{V_1}} \\
\displaystyle + \frac{{\eta \nu }}{{{16\pi }}}\int\limits_{V_1} {\int\limits_{V_2}  {\frac{1}{{{\nu^2}}} \IM \left\{ {{{\V{J}}^\ast}\left( {{\V{r}_1}} \right) \cdot { {\V{J}}\left( {\V{r}_2} \right)}} \right\}\mathcal{G} \left(\boldsymbol{r}_1, \boldsymbol{r}_2\right)\D{V_2}} \D{V_1}} \\
+ \displaystyle\frac{1}{2}\sqrt{\frac{\epsilon}{\mu}} \, \IM \int\limits_\Omega {\left. {\frac{\partial \V{F}}{\partial\omega}} \right|_{I_0}} \cdot \V{F}^\ast\,\mathrm{d} \Omega,
\end{array}
\EE
in which
\BE
\label{TAPcom_EQ8}
\mathcal{G} \left(\boldsymbol{r}_1, \boldsymbol{r}_2\right) = \displaystyle\frac{k^2 \left( \|\boldsymbol{r}_2\|^2 - \|\boldsymbol{r}_1\|^2\right) \mathrm{j}_1 \left(k R\right)}{R},
\EE
with $\mathrm{j}_1 \left(k R\right)$ being the spherical Bessel function of the 1st order. The results of (\ref{TAPcom_EQ7}) are also depicted in Fig.~\ref{fig1} and Fig.~\ref{fig2}. The error in (\ref{TAPcom_EQG1}) observed in Figs.~\ref{fig1},~\ref{fig2} is however not visible in \cite{Geyi_OnStoredEnergiesAndRadiationQ}, the reason most probably being the assumption of an unrealistic current distribution with a purely reactive input impedance at the driving port. In contrast, Figs.~\ref{fig1},~\ref{fig2} consider a rigorous full-wave solution.
\BF
\includegraphics[width=8.9cm]{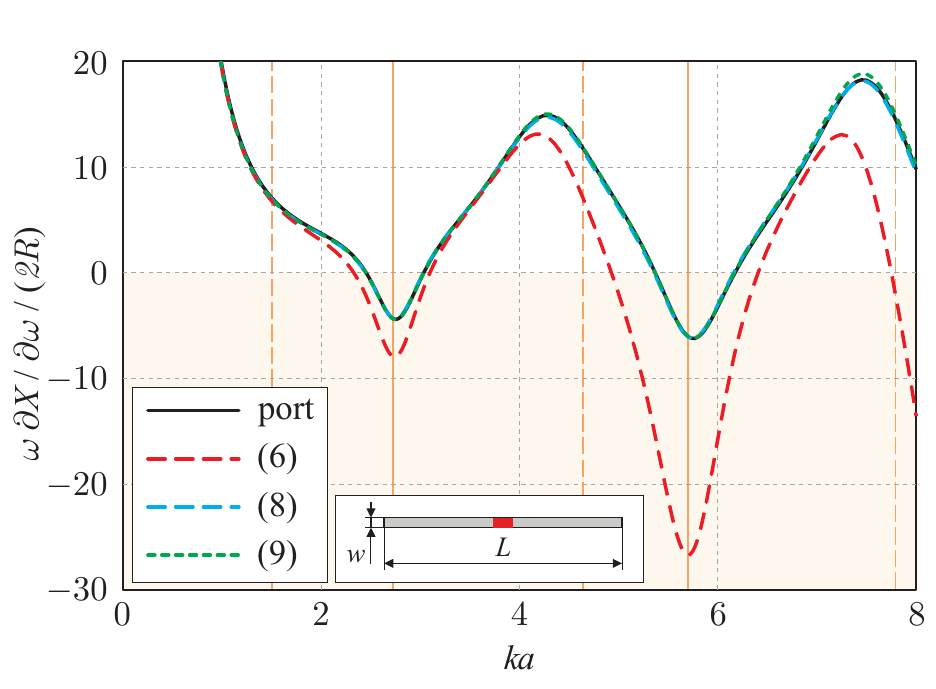}
\caption{The normalized frequency derivative of the input reactance of a thin dipole made of a perfectly electrically conducting strip of width \mbox{$w=L/200$} and of infinitesimal thickness. The dipole is fed in its center by a delta gap source with the input current normalized to $1\,\mathrm{A}$. The quantity $R$ denotes the input resistance of the antenna. The calculation has been performed in the FEKO full-wave integral equation solver \cite{feko}. The frequency derivative of the input reactance is calculated from the port impedance (solid line), from (\ref{TAPcom_EQG1}) (red dashed line), from (\ref{TAPcom_EQ6}) (blue dashed line), and from (\ref{TAPcom_EQ7}) (green dashed line). The dashed orange lines highlight resonances, while the full orange lines highlight antiresonances.}
\label{fig1}
\EF
\BF
\includegraphics[width=8.9cm]{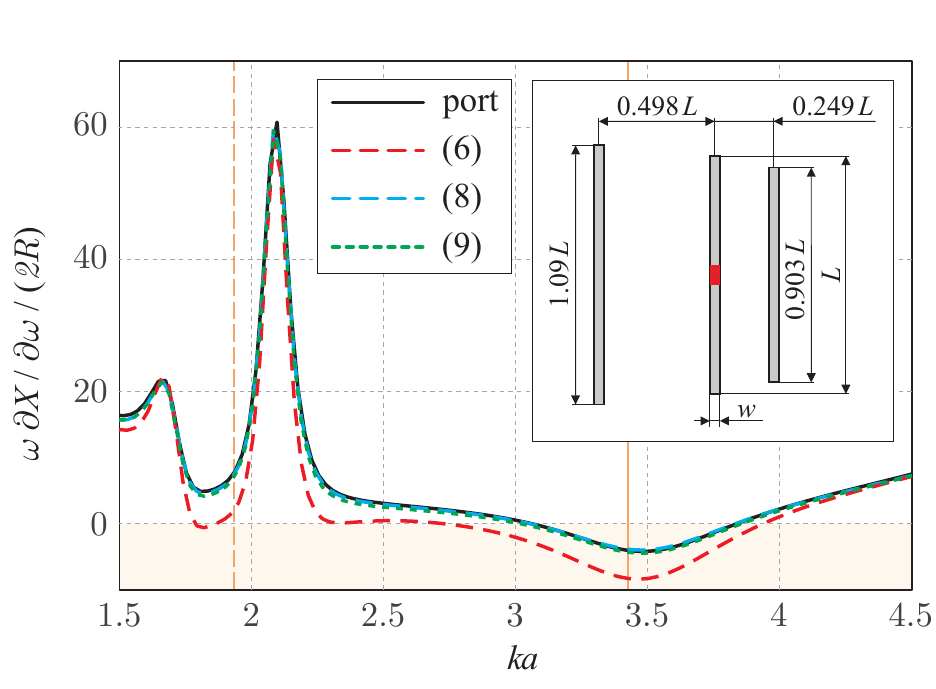}
\caption{The normalized frequency derivative of the input reactance of a thin-strip Yagi-Uda antenna (see the inset). The width of the strip and its thickness, as well as the meaning of depicted curves, are the same as in Fig.~\ref{fig1}.}
\label{fig2}
\EF

As a final comment, it is important to mention that the results of \cite{CapekJelinekHazdraEichler_MeasurableQ,CapekJelinekHazdraEichler_MeasurableQ_Arxive}, for example (\ref{TAPcom_EQ6}), are not mentioned in \cite{Geyi_OnStoredEnergiesAndRadiationQ}, although they address the related problem of the source concept of the quality factor $Q_Z$.


\ifCLASSOPTIONcaptionsoff
  \newpage
\fi

\bibliographystyle{IEEEtran}
\bibliography{references_LIST}

\end{document}